
%
\input phyzzx
\catcode`@=11
%
%
\newtoks\KUNS
\newtoks\HETH
\newtoks\monthyear
\Pubnum={KUNS~\the\KUNS\cr HE(TH)~\the\HETH}
\KUNS={1221}
\HETH={93/11}
\monthyear={September, 1993}
\def\p@bblock{\begingroup \tabskip=\hsize minus \hsize
    \baselineskip=1.5\ht\strutbox \topspace-2\baselineskip
    \halign to\hsize{\strut ##\hfil\tabskip=0pt\crcr
    \the\Pubnum\cr hep-th/9310059\cr \the\monthyear\cr }\endgroup}
\def\bftitlestyle#1{\par\begingroup \titleparagraphs
    \iftwelv@\fourteenpoint\else\twelvepoint\fi
    \noindent {\bf #1}\par\endgroup}
\def\title#1{\vskip\frontpageskip \bftitlestyle{#1} \vskip\headskip}
%
%
\def\acknowledge{\par\penalty-100\medskip \spacecheck\sectionminspace
    \line{\hfil ACKNOWLEDGEMENTS\hfil}\nobreak\vskip\headskip}
%
%

%
\def\journal#1&#2(#3){\begingroup \let\journal=\dummyj@urnal
    \unskip, \sl #1\unskip~\bf\ignorespaces #2\rm
    (\afterassignment\j@ur \count255=#3) \endgroup\ignorespaces}
\def\andjournal#1&#2(#3){\begingroup \let\journal=\dummyj@urnal
    \sl #1\unskip~\bf\ignorespaces #2\rm
    (\afterassignment\j@ur \count255=#3) \endgroup\ignorespaces}
\def\andvol&#1(#2){\begingroup \let\journal=\dummyj@urnal
    \bf\ignorespaces #1\rm
    (\afterassignment\j@ur \count255=#2) \endgroup\ignorespaces}
\def\MPL{Mod.~Phys.~Lett.}
\def\NP{Nucl.~Phys.}
\def\PL{Phys.~Lett.}
\def\PR{Phys.~Rev.}
\catcode`@=12
%

\titlepage

\title{Dynamics of the Cosmological Constant
       \break in Two-Dimensional Universe}

\author{Izawa K.-I.}

\address{Department of Physics, Kyoto University
         \break Kyoto 606, Japan}

\abstract{
We consider a two-dimensional model of gravity
with the cosmological constant as a dynamical variable.
The effective cosmological constant
is derived when the universe has no initial boundary.
It turns out to be extremely small
if the universe is sufficiently large.
}

\endpage


\def\e{\epsilon}
\def\l{\lambda}
\def\m{\mu}
\def\n{\nu}
\def\r{\rho}
\def\L{\Lambda}

\def\i{\int_M \! d^2 \! x \,}
\def\p{\partial}
\def\pf{{1 \over 2\pi}}


\REF\Wei{For a review, S.~Weinberg \journal
         Rev.~Mod.~Phys. &61 (89) 1.}

\REF\Har{J.B.~Hartle and S.W.~Hawking \journal \PR &D28 (83) 2960;
         \nextline S.W.~Hawking \journal \NP &B239 (84) 257.}

\REF\Pol{A.M.~Polyakov \journal \PL &B103 (81) 207;
         \andjournal \MPL &A2 (87) 893; \nextline
         V.G.~Knizhnik, A.M.~Polyakov, and A.B.~Zamolodchikov
         \journal \MPL &A3 (88) 819.}

\REF\Dav{F.~David \journal \MPL &A3 (88) 1651; \nextline
         J.~Distler and H.~Kawai \journal \NP &B321 (89) 509.}

\REF\Dir{P.A.M.~Dirac, {\sl General Theory of Relativity}
         (John Wiley \& Sons, 1975) Chap.~20.}

\REF\Hen{M.~Henneaux and C.~Teitelboim \journal \PL &B222 (89) 195.}

\REF\Hal{A.M.~Polyakov, {\sl Gauge Fields and Strings}
         (Harwood Academic, 1987) Chap.~8; \nextline
         See also
         J.J.~Halliwell and J.B.~Hartle \journal \PR &D41 (90) 1815.}

\sequentialequations

%
{\caps 1. Introduction}

It is a profound mystery about the Universe
that the observational bounds
for the cosmological constant are incredibly small.
This has motivated various ideas
\refmark{\Wei}
on the subject of gravitation and cosmology.

Recent interest in two-dimensional gravity might be largely rooted
in stringy approach to unified theory.
However, we hope that two-dimensional theories of gravity
may also serve as toy models for investigating
qualitative features of realistic gravity in four dimensions.

In this paper, we consider a simple model of two-dimensional gravity
where the cosmological constant appears as a constant of integration.
Namely, the cosmological constant is determined by an initial
condition for a dynamical variable,
whose expectation value
will be computed when the state of the universe
is of the Hartle-Hawking type\rlap.
\refmark{\Har}
The effective cosmological constant turns out to be extremely small
if the universe is sufficiently large.

%
{\caps 2. The Model}

Let us consider the following model Lagrangian
in two dimensions:
$$
  {\cal L} = {\cal L}_c(g_{\m \n}) + \pf (\r \sqrt{}
             + \l \sqrt{} + \l \e^{\m \n} \p_\m A_\n),
 \eqn\MODEL
$$
where ${\cal L}_c(g_{\m \n})$ denotes the effective Lagrangian for
conformal matter with central charge $c$
coupled to gravity\rlap,
\refmark{\Pol, \Dav}
$\r$ is a renormalized cosmological constant,
$\sqrt{}$ represents the invariant volume density
\refmark{\Dir}
in terms of the metric tensor $g_{\m \n}$,
$\l$ is a scalar field
which contributes to the effective cosmological constant,
$A_\m$ is an abelian gauge field, and $\e^{\m \n}$ denotes the
Levi-Civita tensor.

The model \MODEL\ may be regarded as a two-dimensional analogue
of the covariant form of unimodular gravity in four dimensions
given by Henneaux and Teitelboim\rlap.
\refmark{\Hen}
Classically its physical contents are almost the same
as those of the conventional gravity.
A major difference results from
equations of motion
$\p_\m \l = 0$,
which indicate that the effective cosmological
constant appears as a constant of integration.
Thus it is determined by an initial
condition for the universe.

In the following sections,
we will calculate the expectation value of
the observable $\l$ in the case of no-boundary universe,
which involves an initial condition we need.
That is, we adopt as our universe
a hemisphere with $w$ wormholes attached.
Then the desired expectation value will be obtained
as a one-point function $\VEV{\l}$ on a closed Riemann surface
$M$ with $h = 2w$ handles.

%
{\caps 3. Partition Function}

In this section, we estimate
the partition function for the model \MODEL\
exposed in the previous section:
$$
  Z = \int \! {\cal D}g {\cal D}\l {\cal D}A \,e^{-S},
 \eqn\PART
$$
where
$$
  S = \i {\cal L}.
 \eqn\ACTION
$$
We note that integration over the multiplier field $\l$
should be performed
along the direction of the imaginary axis
\refmark{\Hal}
so as to put
the theory \MODEL\ properly in the Euclidean path integral \PART.

Let us first define the zero mode $\l_0$ of the field $\l$
as follows:
$$
  \l = \l_0 + \l_1, \quad \p_\m \l_0 = 0,
 \eqn\ZERO
$$
where $\l_1$ satisfies a condition
$$
  \p_\m \l_1 = 0 \Longleftrightarrow \l_1 = 0.
 \eqn\COND
$$
Then the action \ACTION\ is written as
$$
  S = S_c + \l_0(\pf \! \i \sqrt{} - T)
     + \pf \! \i (\l_1 \sqrt{} + \l_1
      \e^{\m \n} \p_\m A_\n).
 \eqn\EXPAN
$$
Here we have introduced
$$
  S_c = \i ({\cal L}_c + \pf \r \sqrt{}),
   \quad T = - \pf \! \i \e^{\m \n} \p_\m A_\n,
 \eqn\INTR
$$
where $T$ comes out to be a number
which is independent of fluctuation in the field $A_\m$.

The form \EXPAN\ of the action $S$ allows us to perform
successive path integration in \PART\ over
the fields $A_\m$ and $\l_1$ to obtain
$$
  Z = \int {\cal D}g {\cal D}\l_0 \,e^{-S'},
 \eqn\PARTP
$$
where
$$
  S' = S_c + \l_0(\pf \! \i \sqrt{} - T).
 \eqn\ACTP
$$
Further integration over the variable $\l_0$ results in
the expression
$$
  Z = \int {\cal D}g \,e^{-S_c}
      \d (\pf \! \i \sqrt{} - T),
 \eqn\RESUL
$$
which implies that $T$ characterizes the size of the universe.

This expression of the partition function $Z$ makes
its $T$ dependence apparent
\refmark{\Dav}
through scaling behavior:
$$
  Z \sim T^{X} e^{-\r T},
 \eqn\TDEP
$$
where we have defined a constant
$$
  X = {1 \over 12}(h - 1)(25 - c + \sqrt{(25 - c)(1 - c)}) - 1.
 \eqn\SUSEP
$$
Note that the form \TDEP\ is universal in the sense
that it appears independent of the detailed content of
matter-gravity action $S_c$ when the volume $T$ is large.

%
{\caps 4. Cosmological Constant}

Now we proceed to compute
the desired one-point function $\VEV{\l}$.
With the aid of the equations \PARTP\ and \ACTP,
we see
$$
  Z^{-1} {\p \over \p T} Z = \VEV{\l_0} = \VEV{\l},
 \eqn\VEVL
$$
where the last equality follows from the definition \ZERO-\COND.
Thus, by means of \TDEP, we obtain
$$
  \VEV{\l} = {X \over T} - \r.
 \eqn\RESULT
$$

Quantum fluctuation ${\tilde \l}$ is defined by
$$
  \l = \VEV{\l} + {\tilde \l},
 \eqn\DEFF
$$
which satisfies
$\langle {\tilde \l} \rangle = 0$.
Substituting the above expressions into the Lagrangian \MODEL,
we immediately get
$$
  {\cal L} = {\cal L}_c + \pf (\L \sqrt{}
            + {\tilde \l} \sqrt{} + {\tilde \l} \e^{\m \n} \p_\m A_\n
            + \VEV{\l} \e^{\m \n} \p_\m A_\n),
 \eqn\FINAL
$$
where we have written
$$
  \L = {X \over T}.
 \eqn\ECC
$$

As a conceivable interpretation, these results imply that
the effective cosmological constant,
which directly affects the motion of the metric $g_{\m \n}$,
is given by $\L$ with the fluctuation $\tilde \l$
contributing to it no more.
In view of \ECC, we conclude that the effective cosmological constant
is expected to be extremely small when
the universe is sufficiently large.

%
{\caps 5. Discussion}

We have computed the effective cosmological constant \ECC\
in the theory \MODEL\ of two-dimensional gravity
when the state of the universe is of the Hartle-Hawking type.
Two remarks are in order:

\noindent
$i$) Although the value $\L$ is tiny for large $T$,
it turned out to be non-zero.
Observational cosmology suggests that
this feature might be realized in the Universe.

\noindent
$ii$) The effective cosmological constant $\L$ is not
necessarily small when the size of the universe $T$ is not so large.
This might be adequate for inflationary scenarios
which need dominance of the cosmological-constant effect
in an early epoch of the Universe.

It seems interesting to ask whether these features
will be attained in realistic four-dimensional quantum gravity
yet to come.

%
\acknowledge

The author would like to thank H.~Hata, T.~Kugo, and W.~Kummer
for enlightening conversations. He is also grateful to S.~Yahikozawa
for valuable discussions and careful reading of the manuscript.

\refout

\bye